\documentclass[prd,showpacs,nofootinbib,preprintnumbers]{revtex4}%

\usepackage{amssymb,hyperref,amsmath}
\usepackage[dvips]{color}
\usepackage[dvips]{graphicx}
\usepackage{epsfig}

\unitlength=1.2mm \preprint{JLAB-TH-09-992}
\unitlength=1.2mm \preprint{TIFR/TH/09-15}
\begin{document}
\newcommand{\tr}{\mbox{tr}\,}
\newcommand{\Dslash}{{\mathchoice
    {\Dslsh \displaystyle}%
    {\Dslsh \textstyle}%
    {\Dslsh \scriptstyle}%
    {\Dslsh \scriptscriptstyle}}}
\newcommand{\Dslsh}[1]{\ooalign{\(\hfill#1/\hfill\)\crcr\(#1D\)}}
\newcommand{\leftvec}[1]{\vect \leftarrow #1 \,}
\newcommand{\rightvec}[1]{\vect \rightarrow #1 \:}
\renewcommand{\vec}[1]{\vect \rightarrow #1 \:}
\newcommand{\vect}[3]{{\mathchoice
    {\vecto \displaystyle \scriptstyle #1 #2 #3}%
    {\vecto \textstyle \scriptstyle #1 #2 #3}%
    {\vecto \scriptstyle \scriptscriptstyle #1 #2 #3}%
    {\vecto \scriptscriptstyle \scriptscriptstyle #1 #2 #3}}}
\newcommand{\vecto}[5]{\!\stackrel{{}_{{}_{#5#2#3}}}{#1#4}\!}
\newcommand{\vdot}{\!\cdot\!}
\newcommand{\ignore}[1]{}

\bibliographystyle{apsrev}

\title{Bottom-Hadron Mass Splittings from Static-Quark Action on 2+1-Flavor Lattices}

\author{Huey-Wen Lin}
\affiliation{Thomas Jefferson National
Accelerator Facility, Newport News, VA 23606}

\author{Saul D. Cohen}
\affiliation{Thomas Jefferson National
Accelerator Facility, Newport News, VA 23606}

\author{Nilmani Mathur}
\affiliation{Department of Theoretical Physics, Tata Institute of Fundamental Research, Mumbai 400005}

\author{Kostas Orginos}
\affiliation{Thomas Jefferson National Accelerator Facility, Newport News, VA 23606}
\affiliation{Department of Physics, College of William and Mary, Williamsburg, VA 23187-8795}

\date{\today}


\begin{abstract}
We calculate bottom-hadron mass splittings with respect to $B_d$ and $\Lambda_b$ using full QCD with 2+1 flavors of dynamical Kogut-Susskind sea quarks and domain-wall valence quarks along with a static heavy quark. Our lattices have spatial volume of $(2.5\mbox{ fm})^3$ with lattice spacing about 0.124~fm and a range of pion masses as low as 291~MeV. Our results are in agreement with experimental observations and other lattice calculations within our statistical and systematic errors. In particular, we find the mass of the $\Omega_b$ to be consistent with the recent CDF measurement. We also predict the mass for the as yet unobserved $\Xi^\prime_b$ to be 5955(27)~MeV.
\end{abstract}

\maketitle

\section{Introduction}

Recently, there have been exciting developments in heavy-hadron physics, both theoretically and experimentally. Along with discoveries of charmonium states\cite{Barberio:2008fa,Voloshin:2007dx}, considerable progress has been made in studying bottom hadrons. While the $B$ factories, such as Belle and Babar, have investigated the bottom mesons\cite{Barberio:2008fa}, recent experiments at Fermilab have reported the discoveries of a few bottom baryon states. In the summer of 2007, CDF\cite{:2007rw} reported the first observation of the heavy baryons $\Sigma_b$ and $\Sigma_b^*$, and then both D0\cite{:2007ub} and CDF\cite{:2007un} observed the bottom baryon $\Xi_b^-$, breaking a long period of silence following the observation of the $\Lambda_b$ in 1991.
%
The bottom baryon spectrum has become somewhat controversial due to recent results from D0 and CDF. Last summer, D0\cite{Abazov:2008qm} reported a first observation of the doubly strange bottom baryon $\Omega_b^-$ at 6.165(10)(13)~GeV. However, a recent CDF work puts the $\Omega_b^-$ mass at 6.0544(68)(9)~GeV, a difference of 111(12)(14)~MeV. With a discrepancy of 6.2 standard deviations, it appears that the two collaborations cannot both be observing the $\Omega_b$.
%
A theoretical understanding of the bottom baryon spectroscopy from first-principles QCD is crucial and can help disentangle such discrepancies in experiment.
Furthermore, it is anticipated that in the upcoming dedicated bottom physics experiment LHCb at CERN, there will be many more discoveries in the bottom-hadron spectrum. Combining such experiments with improved theoretical understanding, our knowledge about these states will in the near future be significantly enhanced.


Lattice QCD has been successfully computing hadron masses for the last few decades with increasing precision and scope. However, the study of heavy quarks requires special techniques, such as the introduction of an action that will minimize $O(ma)$ errors due to the discretization of space. Such errors are more prominent on coarse lattices due to the heavy quark masses involved. Many existing bottom spectroscopy lattice calculations have been performed in the meson sector.
Ref.~\cite{Li:2008kb} calculated the bottomonium and bottom-light meson spectra using the relativistic heavy-quark action\cite{Christ:2006us,Lin:2006ur} with fermion action parameters determined through nonperturbative tuning to eliminate systematic uncertainties; their ensembles used 2+1 flavors of domain-wall fermions (DWF)\cite{Kaplan:1992bt,Furman:1994ky} with lightest pion mass 275~MeV.
Ref.~\cite{Meinel:2009rd} also used DWF gauge ensembles but with NRQCD\cite{Lepage:1992tx} to calculate bottom-quark quantities.
A more impressive work\cite{Foley:2007ui} done by the TriLat Collaboration used anisotropic 2-flavor Wilson-type lattices with renormalized anisotropy as high as 6 and operators projected into irreducible representations of the lattice cubic group. These techniques allowed them to obtain very clean signals including multiple static-light excited states.
A more complete review of previous studies can be found in the lattice review talks of Refs.~\cite{Kronfeld:2003sd,Wingate:2004xa,Okamoto:2005zg,Onogi:2006km,DellaMorte:2007ny,Gamiz:2008iv} and references therein.

Unlike bottomonium and $B$ mesons, bottom baryons have not received as much attention from lattice QCD. Some pioneering works\cite{Bowler:1996ws,AliKhan:1999yb,Mathur:2002ce} were done using extrapolations with light-fermion actions or the NRQCD action in the quenched approximation, where the fermion loop degrees of freedom are absent. Since it is difficult to estimate the systematic error due to the quenched approximation, high-precision calculations cannot be achieved.
Recently, more dynamical ensembles have become available due to the increase of the computer resources available for numerical research, and since the recent discovery of the double-$b$ baryon, more lattice calculations have emerged.
Ref.~\cite{Lewis:2008fu} calculated single- and double-$b$ baryons with NRQCD action for the bottom quark on isotropic 2+1-flavor clover ensembles, having lightest pion mass around 600~MeV.
Ref.~\cite{Detmold:2008ww} went to a lighter pion mass (275~MeV) using 2+1 flavors of domain-wall fermions but used static-quark action to simulate the bottom quarks. Since the static-quark action tends to result in noisy signals, one needs high statistics to yield numbers comparable with experimental results.
A selection of single-$b$ baryons were calculated using static-quark action on 2-flavor chirally improved lattice Dirac operator at pion masses as light as 350~MeV in Ref.~\cite{Burch:2008qx}.
An ongoing work using staggered and Fermilab fermion actions\cite{ElKhadra:1996mp} on MILC lattices with multiple lattice spacings was presented in Ref.~\cite{Na:2008hz}.

In this work, we report the mass-splittings of the bottom-hadron spectrum using a static-quark action to simulate the bottom quark and using domain-wall fermions for the light valence quarks. We use an extensive set of gauge ensembles of 2+1-flavor staggered-fermion lattices with a range of quark masses resulting in pion masses as light as 290~MeV; the number of available configurations for these ensembles allows us to achieve high statistics. These ensembles have lattice spacing $a=0.124$~fm, and the lattice volume is about $(2.5\mbox{ fm})^3$. It is anticipated that the physical bottom spectrum resembles the spectrum of hadrons with a single static (infinitely massive) quark, where corrections are suppressed by powers of $\Lambda_{\rm QCD}/m_b$, which can be included systematically in heavy-quark effective field theory.

\section{Lattice Setup}\label{sec:setup}
The gauge configurations used in our work were generated by the MILC collaboration using the one-loop tadpole-improved gauge action\cite{Alford:1995hw}, where both $O(a^2)$ and $O(g^2a^2)$ errors are removed. For the fermions in the sea, the asqtad improved Kogut-Susskind action\cite{Orginos:1999cr,Orginos:1998ue,Toussaint:1998sa,Lagae:1998pe,Lepage:1998vj,Orginos:1999kg} is used. This action is the Naik action\cite{Naik:1986bn} ($O(a^2)$-improved Kogut-Susskind action) with the one-link terms smeared such that couplings to gluons with any momentum component equal to $\pi/a$ are set to zero.

For the valence-sector light quarks (up, down and strange), we use the five-dimensional Shamir domain-wall fermion action\cite{Shamir:1993zy,Furman:1994ky}. This action introduces a fifth dimension of extent $L_5$ and a mass parameter $M_5$. In our case, we used $L_5=16$ and $M_5=-1.7$. The physical quark fields $q(\vec x, t)$ reside on the four-dimensional boundaries of the fifth dimension. The left and right chiral components are exponentially separated onto the opposing boundaries, resulting an action with approximate chiral symmetry at finite lattice spacing. The bare-quark mass parameter $(a m^{\rm DWF}_q)$ is introduced as a direct coupling of the boundary chiral components. Hypercubic-smeared (HYP-smeared)\cite{Hasenfratz:2001hp,DeGrand:2002vu,DeGrand:2003in,Durr:2004as} gauge links were used in the domain-wall fermion action to improve chiral symmetry.


Because the valence-quark and sea-quark actions are different, the calculation is inherently partially quenched. That is, the calculation violates unitarity.  Unlike conventional partially quenched calculations, which become unitary when the valence-quark mass is tuned to the sea-quark mass, unitarity cannot be restored by tuning at nonzero lattice spacing. The next-best option is to tune the valence-quark mass in such a way that the resulting pions have the same mass as those made of the sea Kogut-Susskind fermions. In this case unitarity should be restored in the continuum limit, where the $n_f=2$ staggered action has an $SU(8)_L\otimes SU(8)_R\otimes U(1)_V$ chiral symmetry due to the fourfold taste degeneracy of each flavor, and each pion has 15 additional degenerate partners. At finite lattice spacing this symmetry is broken, and the taste multiplets are no longer degenerate but have splittings that are $O(\alpha^2 a^2)$\cite{Orginos:1999cr,Orginos:1998ue,Toussaint:1998sa,Orginos:1999kg,Lee:1999zxa}. The domain-wall fermion mass is tuned to give valence pions that match the Goldstone Kogut-Susskind pion. This is the only Goldstone boson that becomes massless in the chiral limit at nonzero lattice spacing. This choice gives pions that are as light as possible, resulting in better convergence of the chiral perturbation theory (XPT) needed to extrapolate the lattice results to the physical quark mass.  This tuning was performed and used by LHPC collaboration\cite{Renner:2004ck,Edwards:2005kw,Edwards:2006zza,Renner:2007pb,Hagler:2007xi,Edwards:2006qx} as well.

The gauge ensembles used were generated for five pion masses down to 290~MeV with $20^3 \times 64$ lattice volume, corresponding to a physical volume $V=(2.5\mbox{ fm})^3$. The lattice spacing is $a = 0.12406$~fm, as quoted in Ref.~\cite{Aubin:2004wf}, which was determined from $\Upsilon$ spectroscopy. Further information about our lattice actions, parameters and tests of our computational setup can be found in Ref.~\cite{WalkerLoud:2008bp}. The details concerning the ensembles used in this calculation can be found in Table~\ref{tab:numbers}.

\begin{table}
\begin{center}
\begin{tabular}{|c|ccc|}
\hline
Label  	      & $M_\pi$ (GeV) & $N_{\rm conf}$ & $N_{\rm src}$ \\
\hline\hline
m007          &    0.2938(1.1) &            478 &             7 \\
m007$^\prime$ &    0.2938(11) &            429 &            24 \\
m010          &    0.3566(08) &            657 &             6 \\
m020          &    0.4969(06) &            484 &             7 \\
m030          &    0.5982(08) &            564 &             5 \\
\hline
\end{tabular}
\end{center}
\caption{The gauge-ensemble parameters, including the pion mass $M_\pi$ in GeV, the number of gauge configurations $N_{\rm conf}$ and numbers of sources $N_{\rm src}$ used in our calculation. The m007$^\prime$ configurations differ from m007 only in the number of sources used.}\label{tab:numbers}
\end{table}

\subsection{Static-Quark Action}

Starting from the Wilson fermion action on the lattice, we can derive the quark propagator at the static limit, where the quark mass is infinite. In this limit the quark mass is removed, leaving only the forward hopping term in the action:
\begin{equation}
G_Q({\bf x},t;t_0) = \frac{1+\gamma_4}{2}\prod^{t}_{t^\prime=t_0}U_4({\bf x},t^\prime).
\end{equation}
The links $U_4$ that enter this propagator can be any set of gauge covariant paths that connect neighboring points in time. Different choices of paths result in actions that are equivalent up to discretization errors. All of them have the same continuum limit. In our calculation, we tested two choices for the temporal links entering the static quark propagator. We used the simple temporal gauge links of the underlying gauge configuration. In addition, we used HYP-smeared links with the same smearing parameters as those used in the domain-wall action\cite{Hasenfratz:2001hp}. We observed that the HYP-smeared links gave significantly better signals for our correlators; hence, we only report results from that choice.

\subsection{Static-Light Correlation Functions}

At the static-quark limit, the heavy-quark propagator is reduced to a Wilson line. We can construct the static-light hadron correlation functions by contracting the static-quark line with light-quark propagators to create gauge invariant correlation functions. The static-light baryon correlation functions have the form
\begin{equation}
G_{\Gamma}(\vec{x},t;\vec{x}_0,t_0) =
  \langle
    q_f^a(\vec{x},t) \Gamma q_{f^\prime}^b(\vec{x},t)
    \epsilon_{cab}
    P^{cc^\prime}(\vec{x},t;\vec{x}_0,t_0)
    \bar{q}_{f^\prime}^{a^\prime}(\vec{x}_0,t_0) \Gamma \bar{q}_f^{b^\prime}(\vec{x}_0,t_0)
    \epsilon_{c^\prime a^\prime b^\prime}
  \rangle,
\end{equation}
and the static-light meson correlator is
\begin{equation}
M(\vec{x},t;\vec{x}_0,t_0) =
  \langle
    q_f^a(\vec{x},t)
    P^{aa^\prime}(\vec{x},t;\vec{x}_0,t_0)
    \bar{q}_{f^\prime}^{a^\prime}(\vec{x}_0,t_0)
  \rangle,
\end{equation}
where $f \in \{u,d,s\}$ is the light-quark flavor index and $P^{cc^\prime}$ is the Wilson line connecting the source and the sink which are separated by time $t$. The spin matrix $\Gamma$ is either $C\gamma_5$ or $C\gamma_\mu$ for scalar or vector diquarks, respectively.

The quantum numbers (spin $J$, isospin $I$ and strangeness $S$) of the static-light hadronic correlators we construct are listed in Table~\ref{tab:ops}. Since the static quark carries no spin or light-flavor quantum numbers, the numbers in the table correspond to the quantum numbers of the diquark. Since the gamma matrix insertions in the baryon correlators are symmetric ($C\gamma_\mu$) or anti-symmetric ($C\gamma_5$), the corresponding flavor wavefunctions are symmetric (sextet) or antisymmetric (antitriplet), respectively. This flavor symmetry structure is automatically enforced without explicit symmetrization or antisymmetrization of the flavor wavefunctions. In the static limit the hyperfine splittings are exactly zero, so we cannot compute such spin splittings. For the physical bottom quark the hyperfine splittings are less than one percent. When comparing to experiment, we identify our states with the lowest-spin physical particles.

\begin{table}
\begin{center}
\begin{tabular}{|c|ccc|}
\hline
State          & $J$ & $I$ & $S$  \\
\hline \hline
$B_l$          & 1/2 & 1/2 &   0  \\
$B_s$          & 1/2 &   0 & $-1$ \\
\hline
$\Lambda_b$    &   0 &   0 &   0  \\
$\Sigma_b$     &   1 &   1 &   0  \\
$\Xi_b$        &   0 & 1/2 & $-1$ \\
$\Xi_b^\prime$ &   1 & 1/2 & $-1$ \\
$\Omega_b$     &   1 &   0 & $-2$ \\
\hline
\end{tabular}
\end{center}
\caption{\label{tab:ops}The (diquark) quantum numbers of the states studied in this work. The names used are motivated by the flavor content of the states.}
\end{table}

\section{Numerical Results}\label{sec:num}

\subsection{Effective Mass Plots}

The hadron spectrum can be calculated in lattice QCD using two-point hadronic correlation functions
\begin{equation}
C(t_0, t) = \left\langle 0 \left| O(t)\,O^\dagger(t_0) \right| 0 \right\rangle,
\end{equation}
using the creation and annihilation operators $O^\dagger$ and $O$ described in Sec.~\ref{sec:setup}. In order to extract the masses of the states of interest, we take momentum ($=0$ to extract the masses) and spin projections at the source and sink. By inserting a complete set of hadronic eigenstates of the Hamiltonian between the creation and annihilation operators we get
\begin{equation}
C(t) = \sum_n A_n e^{- M_n (t - t_0)}, \label{eq:2pt}
\end{equation}
where $A_n=|\langle 0 | O | n \rangle|^2$ is the overlap factor between the $n^{\rm th}$ eigenstate and the state created by the operator, and $M_n$ is the mass (or energy if considering a multi-particle state) for the $n^{\rm th}$ eigenstate. In addition to differences in the flavor and spin structure of the operators, we also employ smearing on the fermion fields to improve spatial overlap with the hadronic wavefunction. We present results using smeared operators at both source and sink (SS), and also using smearing only at the source with a point sink (SP).

At large time $t$, the states with higher masses become exponentially suppressed, and Eq.~\ref{eq:2pt} becomes dominated by the ground state with mass $M_0$. Thus, we expect the correlator at large time to proportional to $e^{-M_0 t}$. A simple manipulation allows us to cancel the overlap factors and extract the ``effective mass'' at each $t$:
\begin{equation}\label{eq:2pt-meffM}
M_{\rm eff}(t+1/2) = \log(C(t)/C(t+1)).
\end{equation}
For sufficiently large lattice time extent and sufficiently small noise, $M_{\rm eff}(t)$ will asymptotically approach $M_0$ at large time. In practice, there will exist a window in which the effective mass is approximately flat, called the plateau, between the early times which are contaminated by excited states and later times which are dominated by statistical noise. By applying more sophisticated manipulations to the correlators, it is possible to address excited states in the effective mass analytically using single or multiple correlators as input; for more details, see Refs.~\cite{Beane:2009gs,Beane:2009ky,Fleming:2009wb,Fleming:2007zz,Lin:2007iq,Fleming:2004hs}. Since we are only interested in the ground state in this work, Eq.~\ref{eq:2pt-meffM} is sufficient to crosscheck the fitted masses.

\begin{figure}[h]
\includegraphics[width=0.55\textwidth]{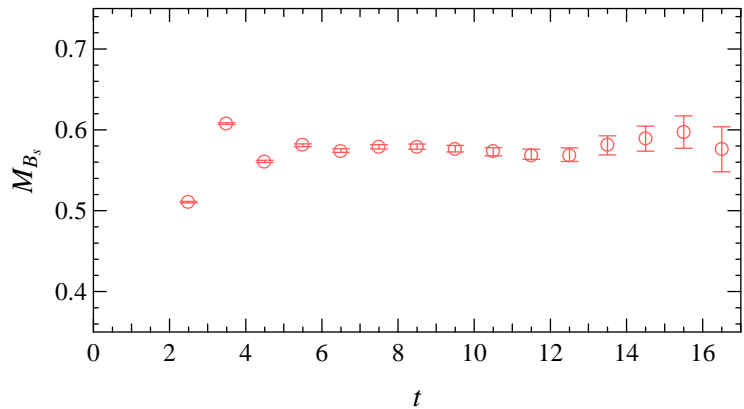}
\caption{\label{fig:effmass}
Effective mass of the lightest $B_s$. Note the oscillation at small time. Both axes are in lattice units.}
\end{figure}

\subsection{Fitting Methodology}

In the (Euclidean) continuum, the two-point correlators follow the form Eq.~\ref{eq:2pt}. We can truncate this expansion at the first term as long as we apply such a single-exponential fit only to the plateau region of each two-point correlator.
However, since we are using the DWF action in the light-valence sector, an additional complication arises from the non-locality of the DWF action in four dimensions. This non-locality manifests as oscillations in the effective mass close to the source, as seen in Fig.~\ref{fig:effmass} in the range $3 \leq t \leq 7$. A phenomenological form for fitting such data was proposed in Ref.~\cite{Renner:2007pb}, employing an oscillating contribution describing the non-local artifacts:
\begin{equation}
C(t) =
  \cdots + A_{\rm osc} \cos(\pi t) e^{-M_{\rm osc} (t-t_0)}, \label{eq:osc}
\end{equation}
where the dots indicate that this is a correction to the continuum form. Since we find empirically that $M_{\rm osc}$ is quite large compared to $M_0$, at large $t$ the oscillations are suppressed and the ground-state mass becomes dominant.

The mass $M_H$ of a hadron with one heavy quark can be expanded in powers of the heavy quark mass:
\begin{equation}
M_H = Z_q m_Q + \Delta + \frac{c}{m_Q} + \cdots\;,
\end{equation}
where $m_Q$ is the mass of the heavy quark and $Z_q$, $\Delta$, and $c$ are arbitrary coefficients. In the static limit the leading term proportional to the heavy-quark mass is eliminated while the sub-leading corrections proportional to $1/m_Q$, although calculable in the heavy-quark effective theory (HQET)\cite{Manohar:2000dt}, are ignored. For this reason, in the static limit only mass splittings between hadrons with the same number of heavy quarks can be easily computed. For two hadrons $B_1$ and $B_2$ the leading term in the ratio of their two-point correlators is
\begin{equation}
(C_{B_1}(t)/C_{B_2}(t)) = A_0^\prime e^{-\Delta M_0(t-t_0)} +
     A_{\rm osc}^\prime \cos(\pi t) e^{-\Delta M_{\rm osc}^\prime (t-t_0)} +
     \cdots,
\label{eq:m-splitting}
\end{equation}
where the dots indicate the effects of the excited-state terms which are suppressed exponentially at large time. The lightest meson ($B_d$) and lightest baryon ($\Lambda_b$) are natural candidates for the denominator of correlator ratios. In two special cases, we try other combinations in order to make more direct comparisons to previous lattice results.

We will compare two kinds of fits: a simple single-exponential and a single-exponential plus the oscillating term. The use of the oscillating term may allow us to extend the fitting range to earlier times, but it comes at the cost of introducing twice as many parameters. We determine the appropriate fit range for our fits using a two-step heuristic: First, calculate the $Q$ value associated with constant fits to the effective mass over all possible ranges. Then, apply the appropriate fit (single-exponential or oscillating) to that range. If extending the range to earlier or later times does not diminish the $Q$ value by more than a factor of 2, extend the range. Continue to extend to the maximum length possible. This heuristic provides a compromise between optimizing the quality of the fit and ensuring that the fit ranges include as much of the data as possible.

\begin{figure}[ht]
\includegraphics[width=0.45\textwidth]{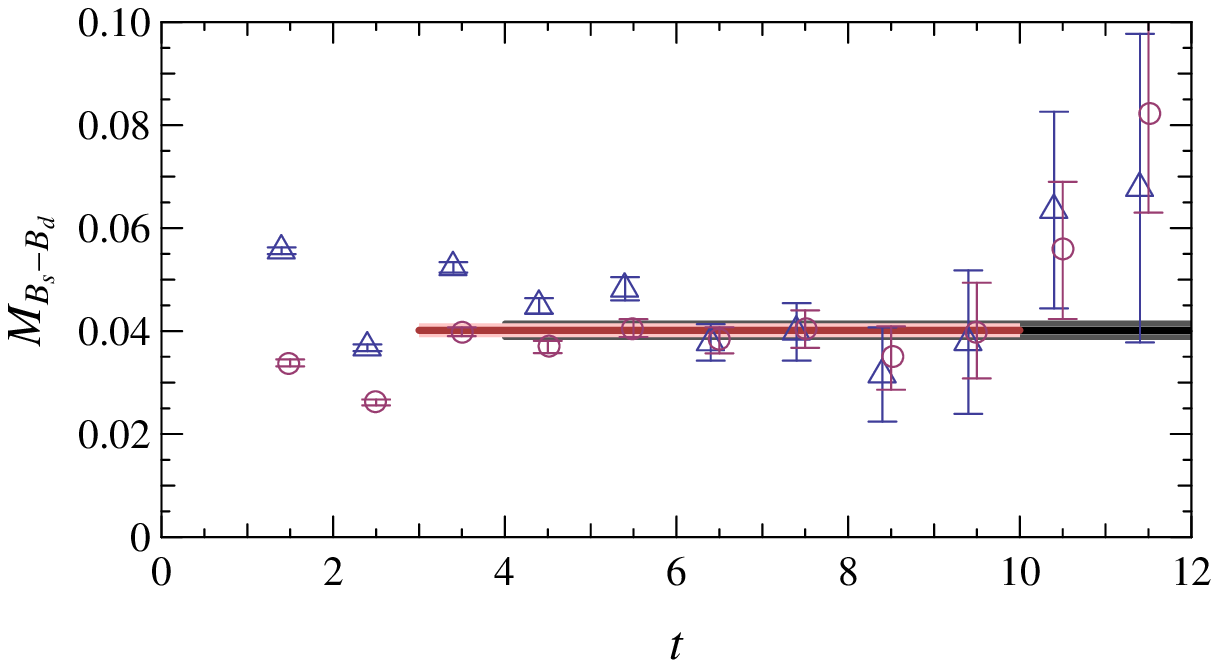}
\includegraphics[width=0.45\textwidth]{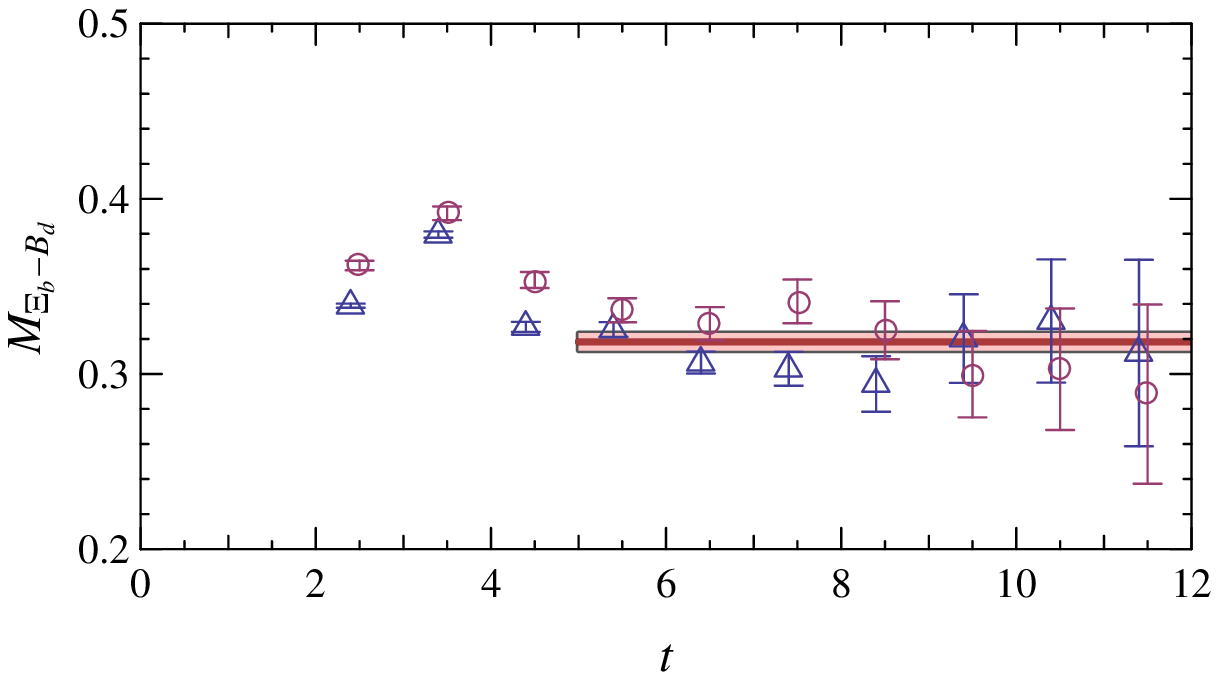}
\includegraphics[width=0.45\textwidth]{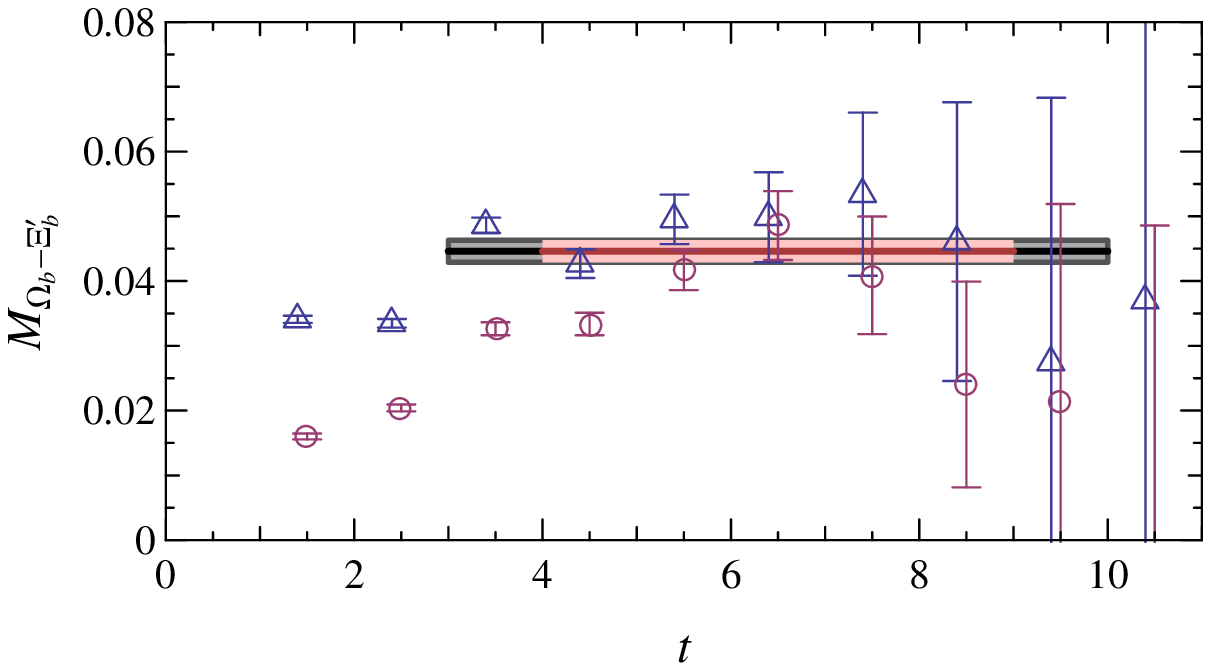}
\includegraphics[width=0.45\textwidth]{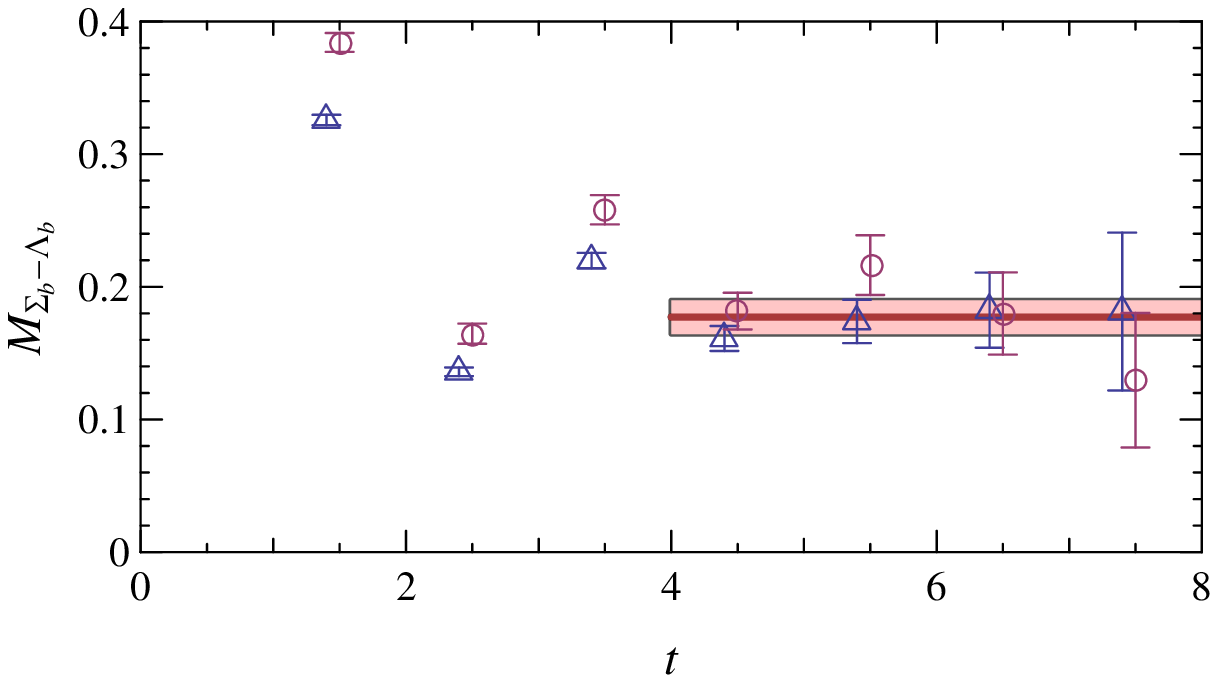}
\caption{\label{fig:SplittingEffmass}
A selection of effective mass splittings showing simultaneous fits to smeared-smeared (dark band) and smeared-point (lighter (red) band) correlators. The (blue) triangles and (red) circles in each panel are the smeared-smeared and smeared-point effective mass splittings. Both axes are in lattice units.}
\end{figure}

Once the appropriate fit range for each correlator ratio is determined, we apply a Levenberg-Marquardt algorithm to minimize $\chi^2$ for each fit form. The procedure is carried out under a single-elimination jackknife to estimate the error in the parameter determinations. We compare the results of fitting the SP two-point correlators using either single-exponential or oscillating fit form in Table~\ref{tab:splittings-SP-comp}.
Since we see good agreement between these two sets of correlators and no uniform improvement by using the more complicated oscillating fit, we adopt the simpler one-state exponential and apply it simultaneously to both SS and SP correlators; the results are shown in Table~\ref{tab:splittings-simultaneous} and a selected fits and its corresponding effective mass plots are also shown in Fig.~\ref{fig:SplittingEffmass}.

\begin{table}
\begin{minipage}[b]{0.45\textwidth}
\begin{tabular}{|c|cccc|}
\hline
$ $ & 0.007 & 0.01 & 0.02 & 0.03 \\
\hline \hline
$B_s-B_d$ & 0.039(01) & 0.043(03) & 0.026(01) & 0.017(01) \\
$\Lambda_b-B_d$ & 0.253(08) & 0.298(08) & 0.300(12) & 0.332(10) \\
$\Xi_b-B_d$ & 0.329(06) & 0.355(07) & 0.332(11) & 0.365(08) \\
$\Sigma_b-B_d$ & 0.438(09) & 0.403(21) & 0.411(13) & 0.437(08) \\
$\Xi_b^{\prime}-B_d$ & 0.456(08) & 0.466(08) & 0.425(15) & 0.455(07) \\
$\Omega_b-B_d$ & 0.484(07) & 0.505(06) & 0.484(05) & 0.473(06) \\
\hline
$\Lambda_b-B_s$ & 0.214(08) & 0.255(10) & 0.256(16) & 0.313(10) \\
$\Xi_b-\Lambda_b$ & 0.099(04) & 0.068(05) & 0.045(01) & 0.033(02) \\
$\Sigma_b-\Lambda_b$ & 0.190(17) & 0.166(09) & 0.106(11) & 0.120(06) \\
$\Xi_b^{\prime}-\Lambda_b$ & 0.218(14) & 0.192(08) & 0.138(10) & 0.144(08) \\
$\Omega_b-\Lambda_b$ & 0.256(13) & 0.224(08) & 0.172(09) & 0.162(08) \\
\hline
\end{tabular}
\end{minipage}
\begin{minipage}[b]{0.45\textwidth}
\begin{tabular}{|c|cccc|}
\hline
$ $ & 0.007 & 0.01 & 0.02 & 0.03 \\
\hline \hline
$B_s-B_d$ & 0.039(02) & 0.043(03) & 0.026(01) & 0.017(01) \\
$\Lambda_b-B_d$ & 0.253(08) & 0.299(08) & 0.282(18) & 0.332(10) \\
$\Xi_b-B_d$ & 0.329(06) & 0.355(07) & 0.318(15) & 0.365(09) \\
$\Sigma_b-B_d$ & 0.438(09) & 0.403(21) & 0.391(20) & 0.437(08) \\
$\Xi_b^{\prime }-B_d$ & 0.456(08) & 0.466(08) & 0.426(15) & 0.455(07) \\
$\Omega_b-B_d$ & 0.485(09) & 0.505(06) & 0.477(07) & 0.473(06) \\
\hline
$\Lambda_b-B_s$ & 0.214(08) & 0.243(13) & 0.256(16) & 0.313(10) \\
$\Xi_b-\Lambda_b$ & 0.099(04) & 0.070(04) & 0.045(02) & 0.033(02) \\
$\Sigma_b-\Lambda_b$ & 0.211(07) & 0.177(07) & 0.106(11) & 0.126(04) \\
$\Xi_b^{\prime }-\Lambda_b$ & 0.234(06) & 0.201(06) & 0.138(10) & 0.141(04) \\
$\Omega_b-\Lambda_b$ & 0.266(06) & 0.231(06) & 0.182(04) & 0.157(04) \\
\hline
\end{tabular}
\end{minipage}
\caption{\label{tab:splittings-SP-comp}The splittings on ensembles with various pion masses from one-exponential (top) and oscillating fits (bottom) to smeared-point correlators. Numbers are given in lattice units.}
\end{table}

\begin{table}
\begin{tabular}{|c|cccc|}
\hline
$ $ & 0.007 & 0.01 & 0.02 & 0.03 \\
\hline \hline
$B_s-B_d$ & 0.040(01) & 0.046(03) & 0.027(01) & 0.018(01) \\
$\Lambda_b-B_d$ & 0.232(06) & 0.282(08) & 0.279(06) & 0.332(10) \\
$\Xi_b-B_d$ & 0.318(05) & 0.342(05) & 0.333(05) & 0.352(05) \\
$\Sigma_b-B_d$ & 0.420(09) & 0.390(16) & 0.386(11) & 0.430(08) \\
$\Xi_b^\prime-B_d$ & 0.441(06) & 0.452(08) & 0.424(08) & 0.448(07) \\
$\Omega_b-B_d$ & 0.477(07) & 0.496(06) & 0.466(06) & 0.467(06) \\
\hline
$\Lambda_b-B_s$ & 0.190(06) & 0.230(07) & 0.252(06) & 0.315(10) \\
$\Xi_b-\Lambda_b$ & 0.101(04) & 0.073(04) & 0.052(02) & 0.034(01) \\
$\Sigma_b-\Lambda_b$ & 0.177(13) & 0.155(06) & 0.121(07) & 0.109(10) \\
$\Xi_b^\prime-\Lambda_b$ & 0.213(09) & 0.175(11) & 0.142(08) & 0.128(09) \\
$\Omega_b-\Lambda_b$ & 0.258(08) & 0.224(08) & 0.179(07) & 0.149(08) \\
\hline
\end{tabular}
\caption{\label{tab:splittings-simultaneous}The splittings on ensembles with various pion masses from one-state exponential fits to smeared-point and smeared-smeared correlators simultaneously. Numbers are given in lattice units.}
\end{table}

\subsection{Chiral Extrapolations}

Having extracted the static-light hadron splitting spectrum for a number of unphysically heavy light-quark masses, we can extrapolate to determine the spectrum at the true light-quark mass. (The strange-quark mass already falls in the range accessible to lattice calculations; we use a single value for it and will assign a systematic error associated with the mismatch between our adopted mass and the physical strange-quark mass.) For sufficiently light masses, it should be possible to employ chiral perturbation theory (XPT) as an extrapolation form. The subject of XPT in the presence of heavy quarks (HXPT) has been addressed in numerous works\cite{Tiburzi:2004kd,Burdman:1992gh,Wise:1992hn,Yan:1992gz,Cho:1992cf,Cho:1992gg,Savage:1995dw}, where Goldstone bosons and photons are the elemental degrees of freedom. HXPT incorporates both the approximate chiral symmetry as well as the heavy-quark symmetry of QCD and have non-analytic terms
in addition to the usual polynomial terms in pion and kaon masses. Such forms could in principle be used to extrapolate lattice data and determine the parameters induced in the theory. In addition, such chiral effective theories can be modified to include the mixed-action corrections in the vacuum\cite{Chen:2009su,Chen:2007ug,Orginos:2007tw}.

However, given the small number of data points available and the large number of parameters in a typical XPT form, a comprehensive chiral extrapolation is not possible. Therefore, we adopt only the leading-order form linear in $M_\pi^2$. Since HXPT\cite{Tiburzi:2004kd} suggests that the next-order non-analytic term can be approximated by a term proportional to $M_\pi^3$, we will also test that form to estimate the systematic error due to uncertainty in the light-quark mass extrapolation.

\begin{figure}[ht]
\includegraphics[width=0.55\textwidth]{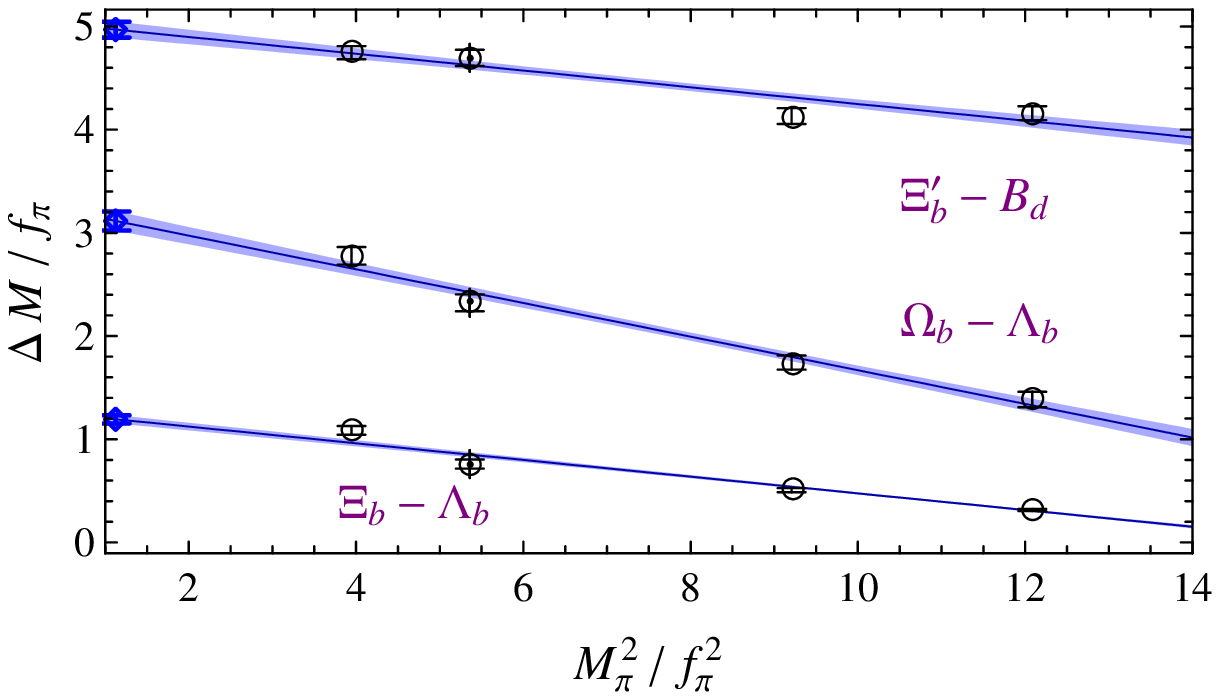}
\includegraphics[width=0.55\textwidth]{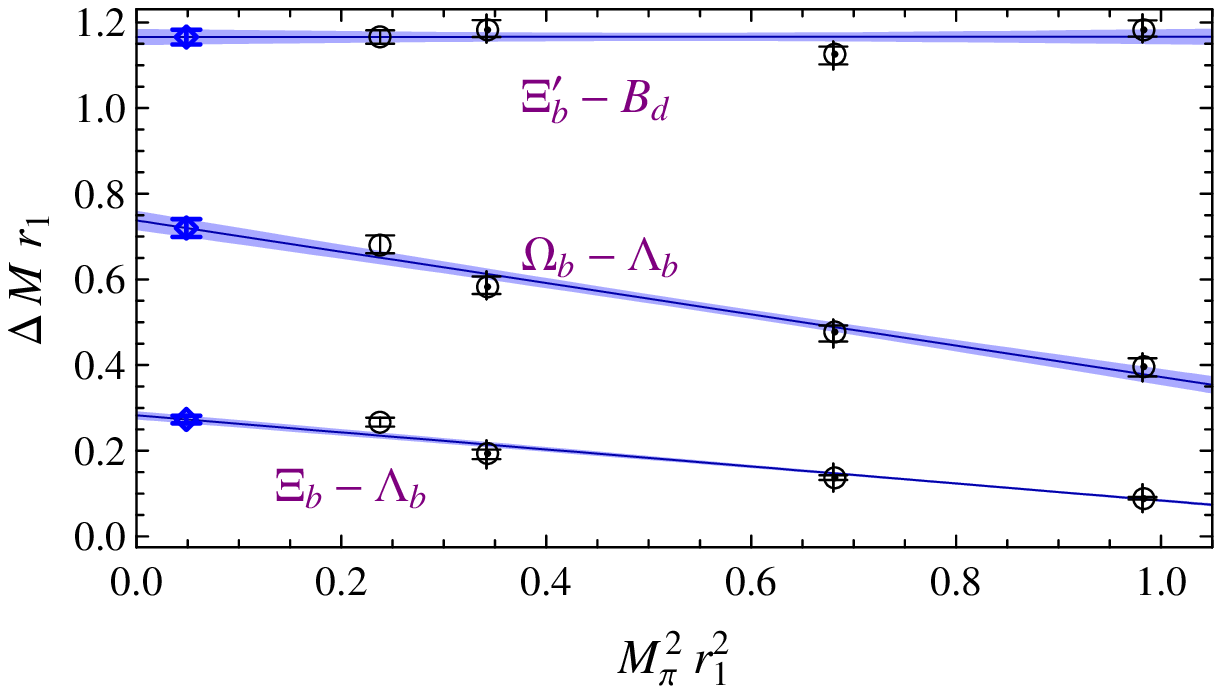}
\includegraphics[width=0.55\textwidth]{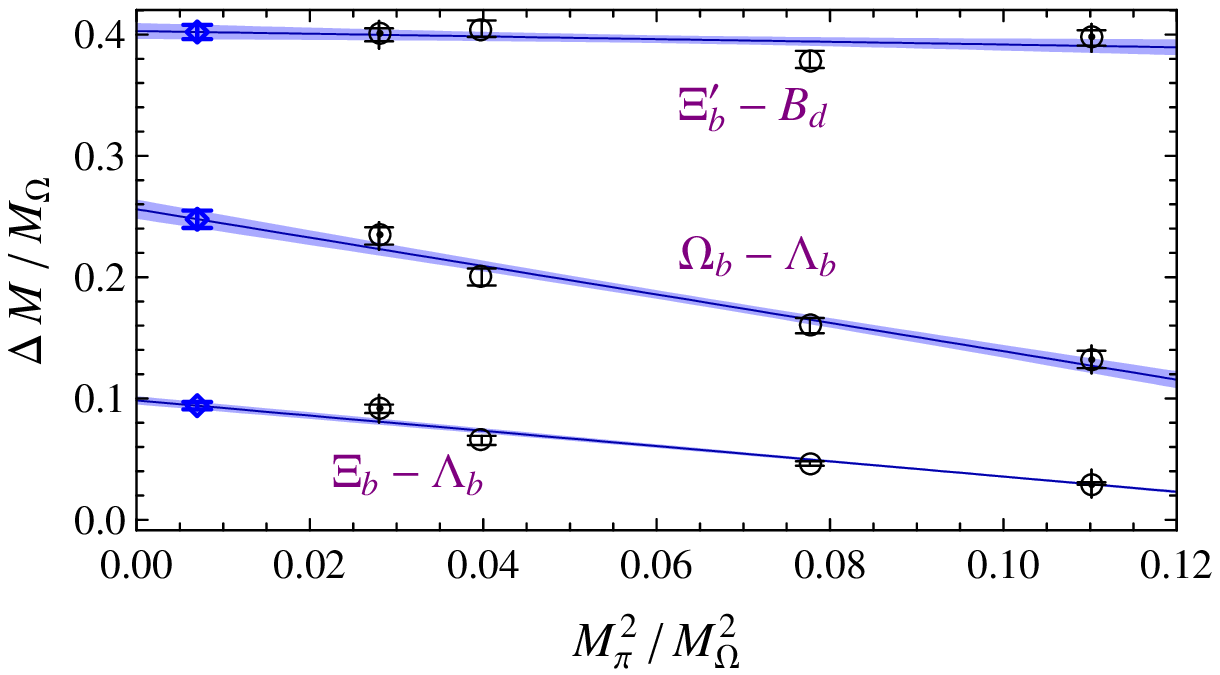}
\caption{\label{fig:extrap}
Selected mass-splitting extrapolations with reference scales $f_\pi$ (top), $r_1$ (middle) and $M_\Omega$ (bottom)}
\end{figure}

Another consideration prompted by the gauge ensembles we have selected is the variation of the lattice spacing between ensembles. Each ensemble uses different light sea-quark mass and different gauge coupling, resulting in variation of the lattice scale. In order to avoid these ambiguities in the scale determination, we have chosen to use dimensionless quantities in the chiral extrapolations. Three different scales, which may be sensitive to different physics, are taken in ratios with the mass splittings and pion mass: the pion decay constant $f_\pi$, the $\Omega$ baryon mass $M_\Omega$ and the Sommer scale of the static-quark potential $r_1$; their values for these ensembles can be found in Refs.~\cite{WalkerLoud:2008bp,Aubin:2004wf}. See Fig.~\ref{fig:extrap} for examples of the extrapolations.

\begin{figure}[ht]
\includegraphics[width=0.65\textwidth]{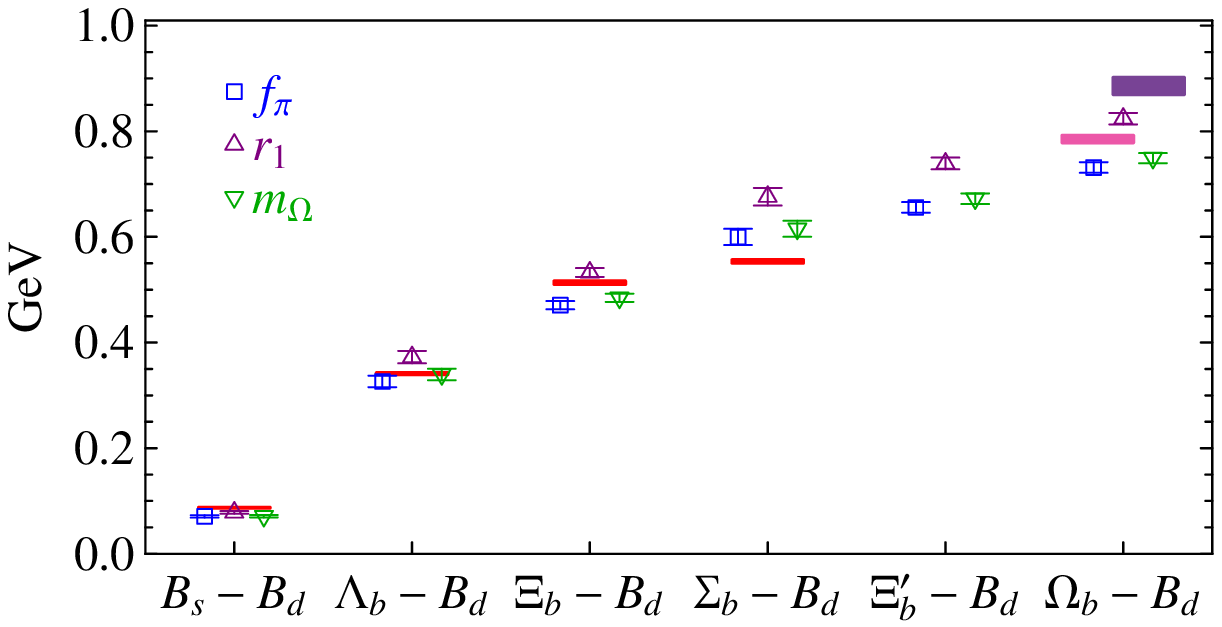}
\includegraphics[width=0.65\textwidth]{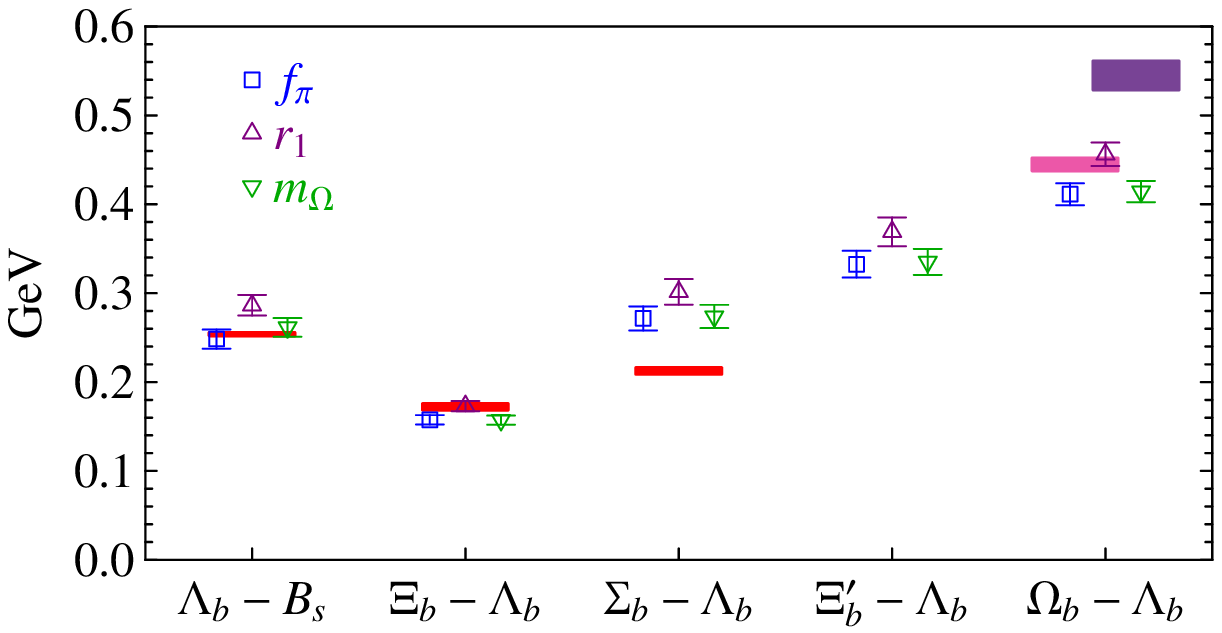}
\caption{\label{fig:mBd-comp}
Comparison of mass-splitting extrapolated values using different reference
scales. Splittings with respect to $B_d$ are on the left; those with respect to
$\Lambda_b$ are on the right. The solid (red) bars indicate the experimental values given in the PDG, where available. For the $\Omega_b$, we show both the D0 result\cite{Abazov:2008qm} (upper-right, purple) and the CDF result\cite{Aaltonen:2009ny} (lower-left, magenta)}
\end{figure}

We summarize the extrapolated mass splittings in Table~\ref{tab:splittings-scaling} and Figs.~\ref{fig:mBd-comp}. We find that there is good consistency between the reference scales $f_\pi$ and $M_\Omega$, but that $r_1$ is slightly away ($\leq 2 \sigma$) from the other two. This discrepancy should probably be interpreted as a systematic error due to the nonzero lattice spacing.
There is also a discretization systematic error due to the single lattice spacing used. One needs to calculate at least two different lattice spacing to be able to extrapolate to the continuum limit. The differences in using methods for setting the lattice spacing are likely to be caused by the lack of such an extrapolation. The discretization systematic can be also estimated by examining the quark action. For example, in our light-valence sector, the action has $O(a^2\Lambda_{\rm QCD}^2)$ error, which is about $4\%$, smaller than the discrepancies we observed here. Therefore, we are not underestimating such a systematic effect here.
Since $M_\Omega$ has smaller dependence on the light-quark mass and is a well measured experimental state, we will use the extrapolated mass splittings normalized by $M_\Omega$ as main estimates of the central values and statistical errors. The discrepancies with other reference scales, we use for estimation of discretization systematic errors.

\begin{table}
\begin{tabular}{|c|ccc|}
\hline
$ $ & $f_\pi$ & $r_1$ & $M_\Omega$ \\
\hline \hline
$B_s-B_d$ &  71.0(2.3)(1.5) &  78.4(2.4)(1.3) &  71.2(2.2)(1.2) \\
$\Lambda_b-B_d$ &  326(11)(11) &  372(12)(9.1) &  340(11)(8.1) \\
$\Xi_b-B_d$ &  470.8(7.9)(8.1) &  532.6(8.5)(7.0) &  484.7(7.7)(6.2) \\
$\Sigma_b-B_d$ &  600(15)(16) &  676(17)(13) &  615(15)(12) \\
$\Xi_b^\prime-B_d$ &  656(10)(12) &  739(11)(10) &  672(10)(9.1) \\
$\Omega_b-B_d$ &  731(10)(12) &  824(11)(10) &  749.2(9.8)(9.0) \\
\hline
$\Lambda_b-B_s$ &  248(11)(11) &  286(12)(9.5) &  261(10)(8.5) \\
$\Xi_b-\Lambda_b$ &  157.5(5.2)(4.3) &  173.0(5.7)(3.9) &  157.2(5.2)(3.4) \\
$\Sigma_b-\Lambda_b$ &  272(14)(17) &  301(14)(14) &  274(13)(13) \\
$\Xi_b^\prime-\Lambda_b$ &  333(15)(14) &  369(16)(12) &  335(15)(10) \\
$\Omega_b-\Lambda_b$ &  411(12)(12) &  456(13)(11) &  414(12)(9.5) \\
\hline
$\Xi_b^\prime-\Sigma_b$ &  61.9(2.7)(2.6) &  69.0(2.9)(2.2) &  62.6(2.6)(2.0) \\
$\Omega_b-\Xi_b^\prime$ &  81.1(2.8)(4.3) &  89.8(3.0)(3.9) &  81.5(2.7)(3.4) \\
\hline
\end{tabular}
\caption{\label{tab:splittings-scaling}The extrapolated splittings using leading-order HXPT with the lattice scale set according to the pion decay constant $f_\pi$, the Sommer scale $r_1$ and the $\Omega$ baryon mass $M_\Omega$. The numbers are given in MeV; first error is statistical and the second is a systematic associated with the mass extrapolation.}
\end{table}

We compare our results with previously published lattice calculations using 2+1 dynamical flavors. For example, Detmold~et~al.\cite{Detmold:2008ww} contains all the mass splittings calculated in this work, and they use a similar lattice spacing. We find good agreement with their numbers, even though we use a different fermion discretization in the sea sector (staggered versus DWF).

\begin{figure}[ht]
\includegraphics[width=0.65\textwidth]{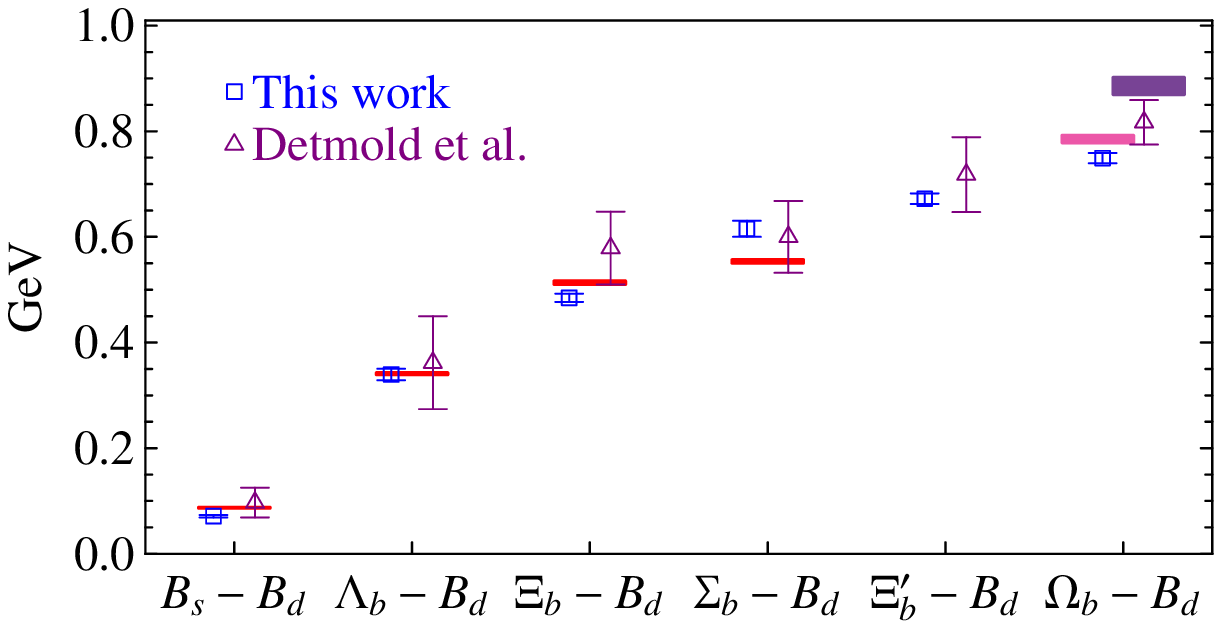}
\includegraphics[width=0.65\textwidth]{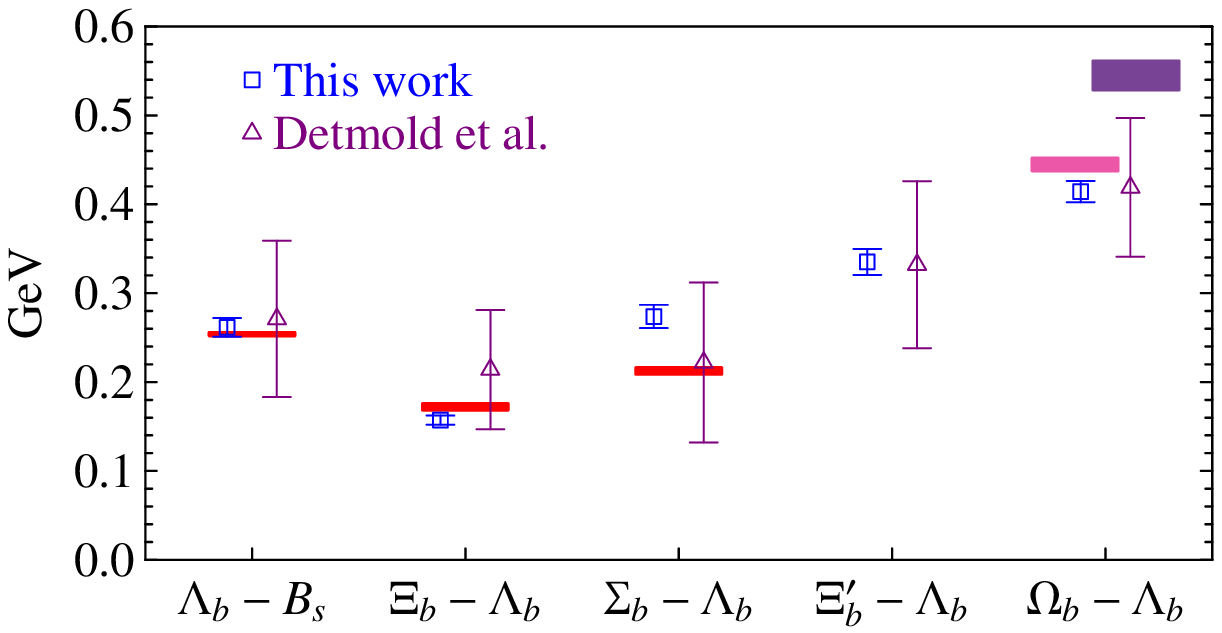}
\caption{\label{fig:ws-mBd-comp}
A comparison of our mass splittings (using $M_\Omega$ reference scale values) with those of Detmold~et~al.\cite{Detmold:2008ww}. The error estimates for both works are statistical only. The solid (red) bars indicate the experimental values given in the PDG, where available. For the $\Omega_b$, we show both the D0 result\cite{Abazov:2008qm} (upper-right, purple) and the CDF result\cite{Aaltonen:2009ny} (lower-left, magenta)
}
\end{figure}

\begin{figure}[ht]
\includegraphics[width=0.65\textwidth]{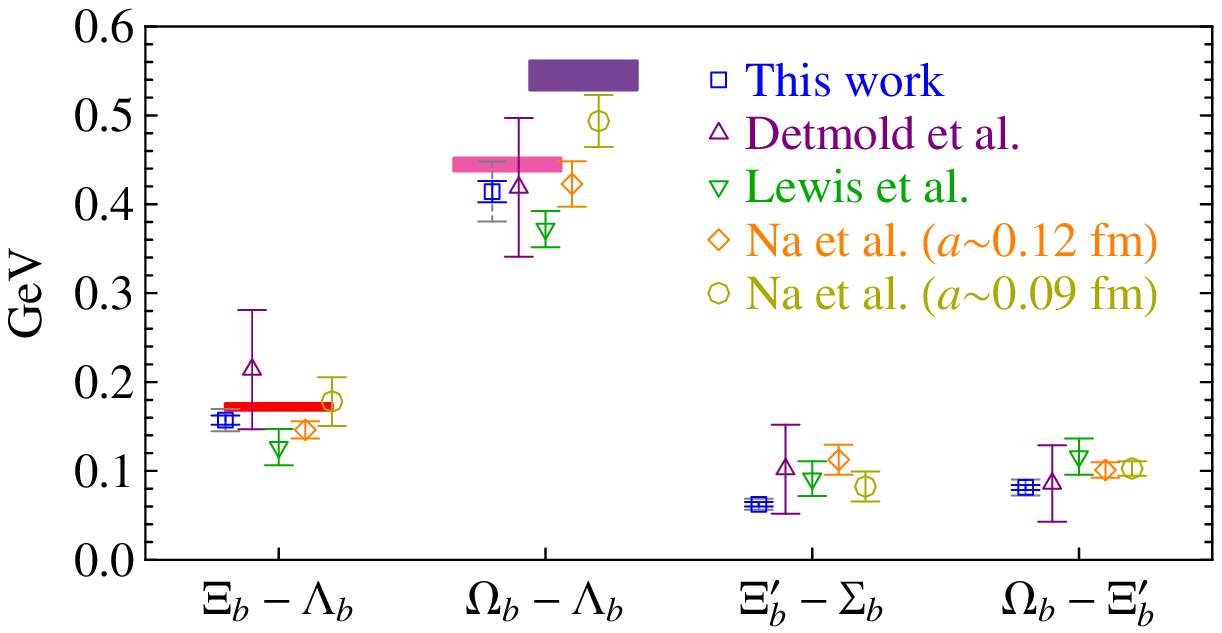}
\caption{\label{fig:all2p1-mixed-comp}
Comparison of mass splittings with all available 2+1-flavor lattice calculations of bottom baryons. The square (blue) points are the points extrapolated using the $\Omega$-mass reference scale; the solid error bars indicate the statistical error, and the dashed bars indicate the total errors (including the estimated systematic ones). The solid (red) bars indicate the experimental values given in the PDG, where available. For the $\Omega_b$, we show both the D0 result\cite{Abazov:2008qm} (upper-right, purple) and the CDF result\cite{Aaltonen:2009ny} (lower-left, magenta).
}
\end{figure}

Unfortunately, other 2+1-flavor works\cite{Na:2008hz,Lewis:2008fu} calculated fewer mass splittings, so we cannot have a comprehensive comparison. Lewis~et~al.\cite{Lewis:2008fu} used the NRQCD action to simulate their bottom quarks on CP-PACS clover lattices; Na~et~al.\cite{Na:2008hz} calculated using the same MILC lattices used in this work but with Fermilab heavy-quark action and staggered light-quark action; they also included another two lattice spacings. All four calculations include the splittings $\Xi_b-\Lambda_b$ and $\Sigma_b-\Lambda_b$; we calculate the additional splittings $\Xi_b^\prime-\Sigma_b$ and $\Omega_b-\Xi_b^\prime$ in order to make direct comparisons. The results are shown in Fig.~\ref{fig:all2p1-mixed-comp}. We again see good agreement amongst all lattice calculations, and mild scaling in the Na~et~al. results. Agreement with experimental values as given by the PDG is fairly good. The discrepancy in the $\Sigma_b-\Lambda_b$ splitting may be due to discretization effects, as suggested by the trend visible in the Na~et~al. results.

We summarize our consensus results for the splittings of the bottom hadron spectrum in Table~\ref{tab:splittings-final}. We adopt the leading-order HXPT extrapolation in terms of the $M_\Omega$ scale as our central values and explicitly give three sources of error: statistical, extrapolation and scale. The extrapolation systematic error is estimated by the discrepancy between the leading-order extrapolation and the result using a form including the next-order $M_\pi^3$ term. These errors are generally about the same size as the statistical error. The error due to scale setting and discretization is estimated using the spread amongst the three scale-setting methods: $f_\pi$, $r_1$ and $M_\Omega$. This systematic is quite large, up to four times the statistical error, and could be resolved by repeating this calculation on finer lattices. The finite-volume correction are negligible for 2.5~fm box, since no such correction is observed in the light-hadron masses\cite{WalkerLoud:2008bp} either. The remaining systematics, such as the effects of the $\Lambda_{\rm QCD}/m_b$ corrections are omitted, since they are substantially smaller  than our main systematics discussed in the text (less than one percent).

These results are in fairly good agreement with experimental results. The largest discrepancies appear for the $\Sigma_b$, although it is only about one sigma away, given the estimated systematics. We give both the D0\cite{Abazov:2008qm} and new CDF\cite{Aaltonen:2009ny} results for the $\Omega_b$ mass. Our result (like most theory predictions) is more consistent with the lower CDF value.

\begin{table}
\begin{tabular}{|c|cc|}
\hline
$ $ & Splitting(Stat.)(Extrap.)(Scale) & Experiment \\
\hline \hline
$B_s-B_d$ &  71.2(2.2)(1.2)(4.2) & 87.1(0.6) \\
$\Lambda_b-B_d$ &  340(11)(8.1)(24) & 341.0(1.6) \\
$\Xi_b-B_d$ &  484.7(7.7)(6.2)(32) & 513(3) \\
$\Sigma_b-B_d$ &  615(15)(12)(40) & 554(3) \\
$\Xi_b^\prime-B_d$ &  672(10)(9.1)(44) & -- \\
$\Omega_b-B_d$ &  749.2(9.8)(9.0)(49) & 786(7)/886(16) \\
\hline
$\Lambda_b-B_s$ &  261(10)(8.5)(19) & 253.9(1.7) \\
$\Xi_b-\Lambda_b$ &  157.2(5.2)(3.4)(9.0) & 172(3) \\
$\Sigma_b-\Lambda_b$ &  274(13)(13)(17) & 213(3) \\
$\Xi_b^\prime-\Lambda_b$ &  335(15)(10)(20) & -- \\
$\Omega_b-\Lambda_b$ &  414(12)(9.5)(25) & 445(7)/545(16) \\
\hline
$\Xi_b^\prime-\Sigma_b$ &  62.6(2.6)(2.0)(3.9) & -- \\
$\Omega_b-\Xi_b^\prime$ &  81.5(2.7)(3.4)(4.9) & -- \\
\hline
\end{tabular}
\caption{\label{tab:splittings-final}Our consensus heavy-hadron splittings in units of MeV. The first error is statistical, the second is a systematic associated with the mass extrapolation and the third is a systematic associated with the setting of the lattice scale.}
\end{table}

\section{Summary and Outlook}\label{sec:end}

In this work, we calculated the mass splittings of bottom baryons using 2+1-flavor staggered-fermion lattice gauge ensembles, domain-wall fermion light valence quarks and static-quark action for the bottom quarks. This is a relatively high-statistics calculation resulting statistical error bars smaller than other recent lattice calculations of baryon mass splittings in the static limit. Our dominant errors are systematic. The most important systematic error is scale setting. We have used three options for converting to physical units and the spread of our final results is an indication of such errors. Such discrepancy is likely to be caused by a nonzero lattice-spacing systematic error in our calculation. A rough estimate of the lattice-spacing systematic error is $4\%$ of the quoted numbers, assuming $O(a^2\Lambda_{QCD}^2)$ scaling violations caused by the light quarks; however, more generous systematic errors are assigned based on the discrepancy from setting the scale using different quantities.
Extrapolations to the physical light-quark mass also introduce errors that should be of the same order of magnitude as typical lattice calculations with the same parameters as ours.  
Finally, $\Lambda_{\rm QCD}/m_b$ corrections (estimated to be $1\%$) have to be included for making direct comparison to experiment. Nonetheless, all the above systematic errors are small, and comparisons with experiment can provide useful conclusions. Our results with our estimates of systematic errors are summarized in Table~\ref{tab:splittings-final}. 

Our results agree well with experiment in the cases of known mass splittings. In the case of $\Omega_b$ our calculation is in agreement with the recent CDF result and several standard deviations away from the D0 result. Such conclusion holds for all other lattice calculations of the $\Omega_b$ mass. Our results for the $\Xi_b^\prime$ mass splittings are a prediction since $\Xi_b^\prime$ has not yet been observed. Using the splitting from the $\Lambda_b$ and summing with the experimental value of that baryon, we predict the mass of the $\Xi_b^\prime$ to be 5955(27)~MeV, where the error given combines in quadrature all our errors.

In the future we plan to extend these calculations using the anisotropic gauge ensembles generated by the Hadron Spectrum Collaboration (HSC)\cite{Lin:2008pr,Edwards:2008ja}. In this case, taking advantage of the anisotropy, a relativistic improved action for the bottom quark may be used, eliminating the need for resorting to the static approximation. As a result, statistical errors are expected to be dramatically improved, and high precision can be achieved, as demonstrated in a recent work\cite{Beane:2009ky} by NPLQCD Collaboration. In this framework, it will also be possible to compute the spin-3/2 to spin-1/2 mass splittings; such effects are inaccessible in the static limit. In addition we expect to have access to lighter pion masses and possibly a second lattice spacing, addressing all sources of systematic error in this calculation. A recent paper by HSC\cite{Peardon:2009gh} demonstrates a new technique, optimized for complicated operators, which may further improve the calculations and in particular allow for the extraction of the excited-state spectrum of the bottom-baryon sector.


\vspace{-0.15cm}
\section*{Acknowledgments}
We thank the NPLQCD collaboration for sharing their propagators: most of the light-quark and all of the strange-quark propagators used in this work; we also thank LHPC for some light-quark propagators. We thank Martin J. Savage for many useful discussions and feedback concerning this work.
These calculations were performed using the Chroma software suite\cite{Edwards:2004sx} on clusters at Jefferson Laboratory using time awarded under the SciDAC Initiative.
KO is supported in part by the Jeffress Memorial Trust grant J-813, DOE OJI grant DE-FG02-07ER41527 and DOE grant DE-FG02-04ER41302.
NM was supported under the grant No. DST-SR/S2/RJN-19/2007.
%
Authored by Jefferson Science Associates, LLC under U.S. DOE Contract No. DE-AC05-06OR23177. The U.S. Government retains a non-exclusive, paid-up, irrevocable, world-wide license to publish or reproduce this manuscript for U.S. Government purposes.

\bibliography{bottom,setup}

\end{document}